\documentstyle[aps,prd,tighten,floats,epsfig,12pt]{revtex}

\def\be{\begin{equation}}
\def\ee{\end{equation}}

\def\re2{\left<r_{i}^{\ 2}\right>}

\begin{document}

\draft
\title{HiX2000: Theory Summary}
\author{A.W. 
Thomas\footnote{athomas@physics.adelaide.edu.au}}

\address{Special Research Centre for the Subatomic Structure of 
Matter and}

\address{Department of Physics and Mathematical Physics, University of
Adelaide, Australia 5005} 

\maketitle
\vspace{-5.5cm}
\begin{flushright}
{\footnotesize Invited talk presented at HiX2000} \\
{\footnotesize Temple University. March 30 --- April 1, 2000} \\
{\footnotesize ADP-00-39/T422 \hspace{2cm}}
\end{flushright}
\vspace{4.5cm}
\begin{abstract}
We summarize the theoretical consensus of the HiX workshop concerning
those measurements which seem most appropriate for inclusion in the
``white paper'' being prepared to justify the proposed 12 GeV upgrade at
Jefferson Lab. The criteria for inclusion are that such measurements
should be decisive, not possible elsewhere and should answer crucial
physics questions with broad implications.
\end{abstract}


\begin{section}{Introduction}

Although a great many innovative and exciting suggestions were made for
physics investigations following the proposed 12 GeV upgrade, it was
eventually necessary to focus on two main themes. We deal with these
in turn below. The first is the possibility of unambiguously determining
the valence $u$ and $d$ quark distributions of the free nucleon in the
region of Bjorken $x$ above 0.4 (and as close as possible to 1). In
particular, while the valence $u$ distribution is relatively well known,
it has recently been realized that the $d$ distribution is very poorly
determined in this region -- at least if one requires that the
determination should be demonstrably model independent. For the
polarized valence distributions, $\Delta u$ and $\Delta d$,
the situation is even worse, with no measurements at all for $x$ above
0.5.

Experiments with $^3$He and $^3$H targets, following the 12 GeV upgrade,
could resolve the longstanding problem concerning the 
valence $d$ distribution for $x$ up to 0.85 -- establishing finally
whether or not the predictions of perturbative QCD are correct. As a
by-product one would complete our knowledge of the EMC effect over the
full range of nuclear mass number. Using polarized proton and 
$^3$He targets one
could also complete the determination of the polarized distributions 
$\Delta u$ and $\Delta d$ over the same region of $x$. The result of
this program would be a definitive picture of the valence structure of
the nucleon, including its spin dependence.

Measurements using polarized proton and $^3$He 
targets at the 12 GeV facility 
would also allow one to definitively establish the twist-3 component of
the structure functions $g_{2p}$ and $g_{2n}$ -- at least for low
moments. This major advance would be made possible by a factor of 10
improvement in statistics over what has hitherto been possible and over
a much greater range of Bjorken $x$: $0.3 \leq x \leq 0.8$. From the
theoretical point of view the twist-3 structure function provides
totally novel information on the internal structure of the nucleon --
information that should be accessible to lattice QCD within the same ten
year time frame.

At a later and more mature stage of development of the facility other
exciting possibilities arise. One could use the knowledge of $u, d,
\Delta u$ and $\Delta d$, at relatively large $x$, to test the validity of
duality in this region and if it works to use it to obtain insight into
these distributions for $x$ as large as 0.95. We could also use this
knowledge to test the utility of semi-inclusive measurements for
determining spin and flavor dependence of parton distributions. If these
tests are successful one could apply these methods to an accurate determination
of the spin and flavor dependence of the nucleon sea in the region $x
> 0.2$. Finally, one could use a tensor polarized deuteron target to
probe completely new structure functions such as $b_1$ for $x$ near and
beyond 1.

\end{section}
\begin{section}{Large-$x$ Valence distributions}\label{sec:Valence}

The distribution of the valence quarks in the nucleon is one of its most
fundamental properties. Because of the relative 4:1 weighting for
electromagnetic probes, the proton structure function primarily
constrains the $u$ distribution in the proton. To determine the $d$
distribution requires a second measurement, traditionally using the
deuteron. While the approximate treatment of the deuteron as a neutron
and a proton at rest, so that $F^n_2 \approx 2F^d_2 - F^p_2$, is not too
bad for $x$ below (say) 0.4, it breaks down badly at large $x$.
Nevertheless it is only in the past few years that it has been realized
that a realistic treatment of the effects of fermi motion and binding in
the deuteron lead to an extracted neutron structure function which is
quite different from that used in all standard parametrizations of 
parton distributions. In particular, a reanalysis of SLAC data by Melnitchouk
and Thomas \cite{MelT}
suggested that the $d/u$ ratio may actually approach the pQCD
prediction of $1/5$ \cite{pQCD}, rather than 0 as $x \rightarrow 1$.

While this analysis is persuasive \cite{recent}, our knowledge of such
fundamental properties should not be model dependent. Fig.~1 shows the
present, appalling state of our knowledge of $d/u$ at large $x$.
\begin{figure}[htb]
\begin{center}
\includegraphics[bb = 35 140 540 710, scale = 0.5]{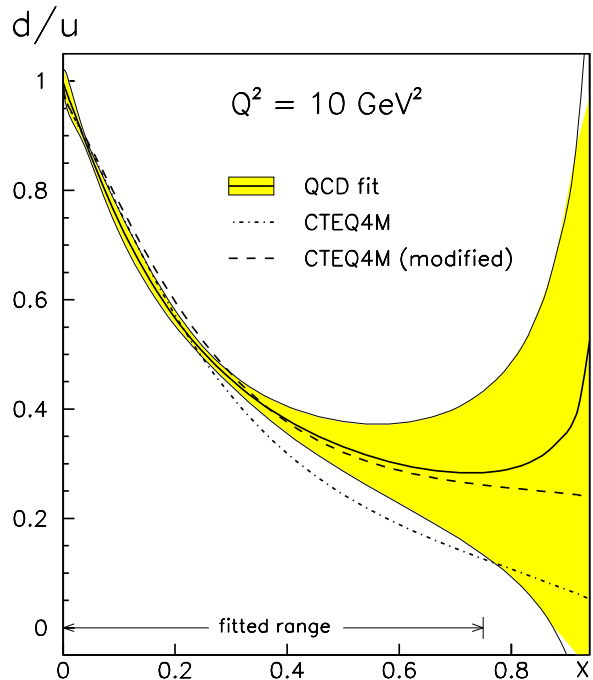}
\caption{Ratio of $(d + \bar d)/(u + \bar u)$ at $Q^2 = 10$ GeV$^2$,
showing the present uncertainty in the region $x > 0.4$ -- from
Ref.~\protect\cite{Botje}.}
\end{center}
\end{figure}
It is clearly vital to
find a facility and a technique that allow us to map out the valence
distributions in a model independent manner over the complete $x$ range
and this problem was addressed by a number of participants at the
workshop including Bosted, Forest, Klein, Leader, Melnitchouk, Meziani, Olness,
Petratos and Scopetta.
It is in this context that the idea of using $^3$He and $^3$H targets is
extremely exciting. The fundamental point is that charge symmetry is a
very good symmetry for strongly interacting systems and these two mirror
nuclei are related by charge symmetry. In the limit of exact charge
symmetry the distribution of protons (the neutron) 
in $^3$He would be the same as that
of neutrons (the proton) in $^3$H -- 
regardless of the size of the EMC effect in
these nuclei! A measurement of the structure functions of these two
nuclei would then allow us to determine $F^n_2/F^p_2$ exactly.

In practice there are electromagnetic corrections to be made and charge
symmetry is slightly broken by the strong interaction. Nevertheless,
exact Faddeev calculations of the structure of the A=3 nuclei allow us
to estimate these corrections quite accurately -- and to test the
dependence of the nuclear structure functions on the two-nucleon
potential used. In order to understand the analysis to be applied to the
data we need to define the ratios:
\be
(R^3{\rm He}) = { F_2^{^3{\rm He}} \over 2 F_2^p + F_2^n }
\hspace{0.5cm}; \hspace{0.5cm} R(^3{\rm H}) =
{ F_2^{^3{\rm H}} \over F_2^p + 2 F_2^n },
\label{eq:1}
\ee
and the super-ratio:
\be
{\cal R} = { R(^3{\rm He}) \over R(^3{\rm H}) }.
\label{eq:2}
\ee
Inverting these expressions directly yields the ratio of the free neutron
to proton structure functions:
\begin{eqnarray}
{ F_2^n \over F_2^p }
&=& { 2 {\cal R} - F_2^{^3{\rm He}}/F_2^{^3{\rm H}}
\over 2 F_2^{^3{\rm He}}/F_2^{^3{\rm H}} - {\cal R} }\ .
\label{eq:3}
\end{eqnarray}

We stress that $F_2^n/F_2^p$ extracted via Eq.(\ref{eq:3}) does not depend
on the size of the EMC effect in $^3$He or $^3$H, but rather on the
{\em ratio} of the EMC effects in $^3$He and $^3$H.
If the neutron and proton distributions in the $A=3$ nuclei are not
dramatically different, one might expect the super-ratio, ${\cal R}$ 
to be approximately one.
To test whether this is indeed the case requires an explicit calculation
of the EMC effect in the $A=3$ system. 
The super-ratio ${\cal R}$, calculated by
Afnan et al. \cite{Afnan}, is shown in
Fig.~2 for the various nuclear model wave functions (PEST, RSC and
Yamaguchi), using the CTEQ parameterization \cite{CTEQ} of parton
distributions at $Q^2=10$~GeV$^2$.
The EMC effects are seen to largely cancel over a large range of $x$,
out to $x \sim 0.85-0.9$, with the deviation from the central value, 
${\cal R} \approx 1.01$, lying within $\pm 1\%$.
Furthermore, the dependence on the nuclear wave function is very weak.
\begin{figure}[htb]
\begin{center}
\epsfig{figure=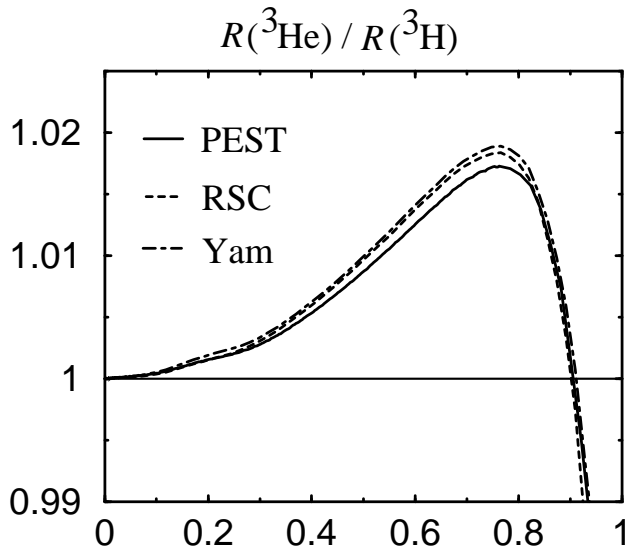,height=8cm}
\caption{Super-ratio of the nuclear EMC ratios for $^3$He and $^3$H for
various nuclear models:
PEST (solid), Reid Soft Core (dashed),
Yamaguchi (dot-dashed) -- from Ref.~\protect\cite{Afnan}.}
\end{center}
\end{figure}

In addition to the dependence on the two-body force used one also needs
to test the sensitivity of ${\cal R}$ to the assumed nucleon structure
functions. The detailed study may be found in Ref.\cite{Afnan}. For our
purposes it is enough to say that while there is some sensitivity one
can adopt a straightforward iterative procedure which is quite rapidly
convergent, so that starting from standard distributions, like CTEQ, one
can quickly converge to a self-consistent set of structure functions and
${\cal R}$. The corollary to this is that having determined 
$F_2^{^3{\rm He}}/F_2^{^3{\rm H}}$ and ${\cal R}$, one will not only pin
down $F_2^n / F_2^p$ but will also determine the EMC effect in $^3$He
and $^3$H, as well as the deuteron, thus completing our experimental
knowledge of the nuclear EMC effect \cite{EMC}. 
This will be a vital assistance in
sorting out the origin of this famous effect \cite{three}, 
especially the possible
change of nucleon structure in a nuclear medium, which is fundamental to
our understanding of QCD itself.

\end{section}
\begin{section}{Spin Dependent Valence Distributions}

As noted earlier, the case for the determination of the spin dependent
structure functions in the region $x > 0.5$ is even clearer -- there is
no data in this region at all! One does not even know, for example,
whether the neutron spin structure function, $g_{1n}$, becomes positive
in this region! This was highlighted, for example, in the presentation
of Leader who made important connections to ``old fashioned'' analysing
power data ($A^{\uparrow} + B \rightarrow C + X$) \cite{Leader}. 
It is remarkable that the focus on the ``spin crisis''
has diverted so much effort to small $x$ that nothing is known in the
large $x$ region. While there is room for more work, especially on
exchange current corrections (mesons, $\Delta$'s, etc.), it seems that
one can determine $g_{1n}$ relatively well from measurements with a
polarized $^3$He target \cite{Liuti}, while accurate data on the proton may 
be simplified by the development of a better $\vec{p}$ target.

It is clear that Jefferson Lab, with the 12 GeV upgrade, can provide a
definitive picture of the spin dependent valence structure functions in
the large $x$ region, thus completing one of the fundamental tasks of
the international program in deep inelastic scattering. On the
theoretical side, apart from the comparison with QCD inspired models
\cite{DIS}, one can expect these distributions to be accessible to
lattice QCD over the same time frame \cite{latt}.

\end{section}
\begin{section}{Twist-3 Structure Functions}

A compelling case was presented by Ji, Bosted, Averett, Meziani and
others that following the proposed upgrade, Jefferson Lab could
determine $g_2$ for both the proton and neutron, with an order of
magnitude improvement in statistical accuracy in the range $x \in (0.3,0.8)$.
This would enable one to remove the trivial twist-2
contribution and unambiguously isolate the twist-3 piece. As this
involves {\it totally novel} information on the internal structure of
the target, it should provide extremely important new information on 
the internal structure of hadrons \cite{Ji}. For example, the second moment of
the twist-3 part of $g_2$, $d_2$, involves the matrix elements
$\langle ps | \psi^\dagger \vec{B} \psi | ps \rangle$ and 
$\langle ps | \psi^\dagger \vec{\alpha} \times \vec{B} \psi | ps \rangle$, 
where $\vec{B}$ is the color magnetic field inside the hadron. This will
be the first indication of the correlation of the quark and gluon fields
inside a hadron.

\end{section}
\begin{section}{Conclusion}

The consensus of the participants at this workshop were that, following
an upgrade to 12 GeV, Jefferson Lab would be able to provide important
and definitive answers to our questions concerning the spin and flavor
distributions of the valence quarks in the nucleon. It would also be
able to provide the first unambiguous information on the twist-3
structure functions of the nucleon, that is the correlations between
quarks and gluons in the proton. This represents an outstanding 
physics program.

As a side benefit of the determination of the valence $d$ distribution
we would also have accurate measurements of the EMC effect in $^3$He,
$^3$H and the deuteron for the first time, thus allowing a complete
study of the nuclear EMC effect and the possible change of nucleon
structure in medium.

{}Following these top priority investigations there are numerous other
important studies to be made. It is vital to explore the validity of
Bloom Gilman scaling \cite{BG} for spin 
structure functions \cite{WM}. As we have already
seen from Jefferson Lab data, scaling may well work where one has no
theoretical reason to justify it (yet). If it can be shown to work one
could use the technique to study a host of quantities at large $x$,
notably $\bar d - \bar u$, $s - \bar s$, $\Delta s$, etc.  As explained
in detail by Mulders, Leader and Ent \cite{Mulders}, 
semi-inclusive measurements allow us 
probe a large number of new observables. As emphasised by Mitchell, 
one can also use a tensor
polarized target to determine the new spin structure function $b_1$,
which is expected to be significant at large $x$. One could investigate
the spin dependence of the EMC effect at large $x$ \cite{steffens} using
$^3\vec{\rm He}$ and $\vec{\rm D}$ targets. As emphasised by Brodsky 
and Liuti, one can investigate quark
and gluon correlations in the nucleon by measuring higher twist
structure functions. As discussed by Kumar, parity violation can also be
exploited to study parton distributions.

In summary, there is an urgent physics case for the 12 GeV upgrade in
order to answer vital and topical questions which go to the heart of our
understanding of strongly interacting systems. 

\end{section}

\section*{Acknowledgement}
I would like to thank Wally Melnitchouk and Zein-Eddine Meziani for
their invitation to participate in this workshop and the hospitality
during what proved to be a very stimulating meeting.
This work was supported by the Australian Research Council and the
University of Adelaide.  



\begin{references}
%
\bibitem{MelT}
W.~Melnitchouk and A.~W.~Thomas,
Phys.\ Lett.\  {\bf B377}, 11 (1996)
[nucl-th/9602038].
%
\bibitem{pQCD}
G.~R.~Farrar and D.~R.~Jackson,
Phys.\ Rev.\ Lett.\  {\bf 35}, 1416 (1975);
%
S.~J.~Brodsky, M.~Burkardt and I.~Schmidt,
Nucl.\ Phys.\  {\bf B441}, 197 (1995)
[hep-ph/9401328].
%
\bibitem{recent}
U.~K.~Yang and A.~Bodek,
Phys.\ Rev.\ Lett.\  {\bf 82}, 2467 (1999)
[hep-ph/9809480].
%
\bibitem{Botje}
M.~Botje,
Eur.\ Phys.\ J.\  {\bf C14}, 285 (2000)
[hep-ph/9912439].
%
\bibitem{Afnan}
I.~R.~Afnan, F.~Bissey, J.~Gomez, A.~T.~Katramatou, W.~Melnitchouk,
G.~G.~Petratos and A.~W.~Thomas,
nucl-th/0006003.
%
\bibitem{CTEQ}
H.~L.~Lai {\it et al.}  [CTEQ Collaboration],
Eur.\ Phys.\ J.\  {\bf C12}, 375 (2000)
[hep-ph/9903282].
%
\bibitem{EMC}
J.~J.~Aubert {\it et al.}  [European Muon Collaboration],
Phys.\ Lett.\  {\bf B123}, 275 (1983);
%
D.~F.~Geesaman, K.~Saito and A.~W.~Thomas,
Ann.\ Rev.\ Nucl.\ Part.\ Sci.\  {\bf 45}, 337 (1995);
%
M.~Arneodo,
Phys.\ Rept.\  {\bf 240}, 301 (1994).
%
\bibitem{three}
W.~Melnitchouk, I.~R.~Afnan, F.~Bissey and A.~W.~Thomas,
Phys.\ Rev.\ Lett.\  {\bf 84}, 5455 (2000)
[hep-ex/9912001].
%
\bibitem{Liuti}
C.~Ciofi degli Atti and S.~Liuti,
Phys.\ Rev.\  {\bf C41}, 1100 (1990).
%
\bibitem{DIS}
A.~W.~Schreiber, P.~J.~Mulders, A.~I.~Signal and A.~W.~Thomas,
Phys.\ Rev.\  {\bf D45}, 3069 (1992).
%
\bibitem{latt}
C.~Best {\it et al.},
hep-ph/9706502.
%
\bibitem{Leader}
M.~Boglione and E.~Leader,
hep-ph/0005092.
%
\bibitem{Ji}
X.~Ji and J.~Osborne,
Eur.\ Phys.\ J.\  {\bf C9}, 487 (1999)
[hep-ph/9902393].
%
\bibitem{BG}
E.D. Bloom and F. J. Gilman, Phys.\ Rev.\ Lett.\ {\bf 16},
1140 (1970).
%
\bibitem{WM}
W.~Melnitchouk,
hep-ph/9909463.
%
\bibitem{Mulders}
A.~Bacchetta and P.~J.~Mulders,
hep-ph/0006131.
%
\bibitem{steffens}
F.~M.~Steffens, K.~Tsushima, A.~W.~Thomas and K.~Saito,
Phys.\ Lett.\  {\bf B447}, 233 (1999)
[nucl-th/9810018].
%
\end{references}
\end{document}